\documentclass[amsmath,amssymb,superscriptaddress,balancelastpage,prx,twocolumn]{revtex4}

\usepackage{nicefrac}
\usepackage{xfrac}
\usepackage{hyperref}
\hypersetup{colorlinks,linkcolor=blue,urlcolor=blue,citecolor=blue}
\usepackage{siunitx}
\usepackage{xfrac}
\usepackage{natbib}
\usepackage{graphicx}
\usepackage{varioref}
\usepackage{xr-hyper}
\usepackage{xcolor}
\usepackage{colortbl}
\usepackage{hyperref}
\usepackage{ulem}
\usepackage{braket}
\usepackage{soul}

\newcommand{\hili}[1]{{\textcolor[rgb]{0,0,0}{#1}}}
\newcommand{\hilia}[1]{{\textcolor[rgb]{0 ,0,0}{#1}}}

\newcommand{\Fe}{$^{57}$Fe }
\newcommand{\Se}{$^{77}$Se }
\newcommand{\Tnem}{$ T_{nem} $}
\newcommand{\Tc}{$ T_{c} $}

\newcommand{\dxy}{$3d_{xy} ${ }}
\newcommand{\dxz}{$3d_{xz} ${ }}
\newcommand{\dyz}{$3d_{yz} ${ }}

\usepackage{multirow}
\usepackage{bbold}
\usepackage{subfigure}
\usepackage{color}
\usepackage{mathrsfs}
\usepackage{hyperref}
\begin{document}

\title{Spin-orbital-intertwined nematic state in FeSe}

\author{J.~Li}
\affiliation{Hefei National Laboratory for Physical Sciences at the Microscale and Department of Physics, University of Science and Technology of China, Hefei, Anhui 230026, China}

\author{B.~Lei}
\affiliation{Hefei National Laboratory for Physical Sciences at the Microscale and Department of Physics, University of Science and Technology of China, Hefei, Anhui 230026, China}

\author{D.~Zhao}
\affiliation{Hefei National Laboratory for Physical Sciences at the Microscale and Department of Physics, University of Science and Technology of China, Hefei, Anhui 230026, China}

\author{L. P.~Nie}
\affiliation{Hefei National Laboratory for Physical Sciences at the Microscale and Department of Physics, University of Science and Technology of China, Hefei, Anhui 230026, China}

\author{D. W.~Song}
\affiliation{Hefei National Laboratory for Physical Sciences at the Microscale and Department of Physics, University of Science and Technology of China, Hefei, Anhui 230026, China}

\author{L. X.~Zheng}
\affiliation{Hefei National Laboratory for Physical Sciences at the Microscale and Department of Physics, University of Science and Technology of China, Hefei, Anhui 230026, China}

\author{S. J.~Li}
\affiliation{Hefei National Laboratory for Physical Sciences at the Microscale and Department of Physics, University of Science and Technology of China, Hefei, Anhui 230026, China}

\author{B. L.~Kang}
\affiliation{Hefei National Laboratory for Physical Sciences at the Microscale and Department of Physics, University of Science and Technology of China, Hefei, Anhui 230026, China}

\author{X. G.~Luo}
\affiliation{Hefei National Laboratory for Physical Sciences at the Microscale and Department of Physics, University of Science and Technology of China, Hefei, Anhui 230026, China}
\affiliation{Key Laboratory of Strongly-coupled Quantum Matter Physics, Chinese Academy of Sciences, University of Science and Technology of China, Hefei, Anhui 230026, China}
\affiliation{CAS Center for Excellence in Superconducting Electronics (CENSE), Shanghai 200050, China}
\affiliation{CAS Center for Excellence in Quantum Information and Quantum Physics, Hefei, Anhui 230026, China}
\affiliation{Collaborative Innovation Center of Advanced Microstructures, Nanjing University, Nanjing 210093, China}

\author{T.~Wu}\email{wutao@ustc.edu.cn}
\affiliation{Hefei National Laboratory for Physical Sciences at the Microscale and Department of Physics, University of Science and Technology of China, Hefei, Anhui 230026, China}
\affiliation{Key Laboratory of Strongly-coupled Quantum Matter Physics, Chinese Academy of Sciences, University of Science and Technology of China, Hefei, Anhui 230026, China}
\affiliation{CAS Center for Excellence in Superconducting Electronics (CENSE), Shanghai 200050, China}
\affiliation{CAS Center for Excellence in Quantum Information and Quantum Physics, Hefei, Anhui 230026, China}
\affiliation{Collaborative Innovation Center of Advanced Microstructures, Nanjing University, Nanjing 210093, China}

\author{X. H.~Chen}\email{chenxh@ustc.edu.cn}
\affiliation{Hefei National Laboratory for Physical Sciences at the Microscale and Department of Physics, University of Science and Technology of China, Hefei, Anhui 230026, China}
\affiliation{Key Laboratory of Strongly-coupled Quantum Matter Physics, Chinese Academy of Sciences, University of Science and Technology of China, Hefei, Anhui 230026, China}
\affiliation{CAS Center for Excellence in Superconducting Electronics (CENSE), Shanghai 200050, China}
\affiliation{CAS Center for Excellence in Quantum Information and Quantum Physics, Hefei, Anhui 230026, China}
\affiliation{Collaborative Innovation Center of Advanced Microstructures, Nanjing University, Nanjing 210093, China}

\begin{abstract}
$\newline$

{The importance of the spin-orbit coupling (SOC) effect in Fe-based superconductors (FeSCs) has recently been under hot debate. Considering the Hund's coupling-induced electronic correlation, the understanding of the role of SOC in FeSCs is not trivial and is still elusive. Here, through a comprehensive study of  $^{77}$Se and $^{57}$Fe nuclear magnetic resonance, a nontrivial SOC effect is revealed in the nematic state of FeSe. First, the orbital-dependent spin susceptibility, determined by the anisotropy of the $^{57}$Fe Knight shift, indicates a predominant role from the 3\textit{d$_{xy}$} orbital, which suggests the coexistence of local and itinerant spin degrees of freedom (d.o.f.) in the FeSe. Then, we reconfirm that the orbital reconstruction below the nematic transition temperature (\textit{T$_{nem}$} $\sim$  90 K) happens not only on the 3\textit{d$_{xz}$} and 3\textit{d$_{yz}$} orbitals but also on the 3\textit{d$_{xy}$} orbital, which is beyond a trivial ferro-orbital order picture. Moreover, our results also indicate the development of a coherent coupling between the local and itinerant spin d.o.f. below \textit{T$_{nem}$}, which is ascribed to a Hund's coupling-induced electronic crossover on the \dxy orbital. \hilia{Finally, due to a nontrivial SOC effect, sizable in-plane anisotropy of the spin susceptibility emerges in the nematic state, suggesting a spin-orbital-intertwined nematicity rather than simply spin- or orbital-driven nematicity}. The present work not only reveals a nontrivial SOC effect in the nematic state but also sheds light on the mechanism of nematic transition in FeSe.}
\end{abstract}

\maketitle
\begin{center}
    \textbf{I. INTRODUCTION}
\end{center}
Electronic correlation is widely accepted as a predominant physical origin for the production of the various quantum matter states in high-temperature superconductors (HTS), including cuprate superconductors ~\cite{Lee2006} and Fe-based superconductors (FeSCs)~\cite{Hirschfeld2011,Si2016,Fernandes2017}. In contrast, spin-orbit coupling (SOC) due to the relativistic motion of electrons has always been considered insignificant in these HTS families. However, a renaissance on the SOC in FeSCs is greatly promoted by the recent efforts to search for the influence of SOC on physical properties~\cite{Wu2008, Fernandes2017,Christensen2015,Wang2015,Wu2016,Xu2016,Klug2018,Christensen2018,Borisenko2016,Ma2016,Hu2017,Waber2017,Kushnirenko2018,Liu2018,Day2018,Zhang2018,Wang2018,Liu2018-2,Zhang2019}. In general, SOC can intertwine spin and orbital degrees of freedom (d.o.f.) in solids and induce intriguing physical properties such as nontrivial topological band structure~\cite{Hasan2010,Qi2011,Jungwirth2012,Pesin2010} and exotic superconducting pairing~\cite{Gor'kov2001,Lu2015,Xi2016}. Recently, a nontrivial topological band structure has been successfully verified in iron chalcogenide superconductors~\cite{Zhang2018,Wang2018,Liu2018-2} that confirms the importance of SOC on the band structure in FeSCs. A natural follow-up question is how to understand the SOC effect on the various quantum matter states, such as nematic and superconducting states, due to electronic correlation in FeSCs ~\cite{Ma2016,Hu2017,Waber2017,Kushnirenko2018,Liu2018,Day2018}.

\begin{figure}[h!]
	\centering
	\includegraphics[width=0.48\textwidth]{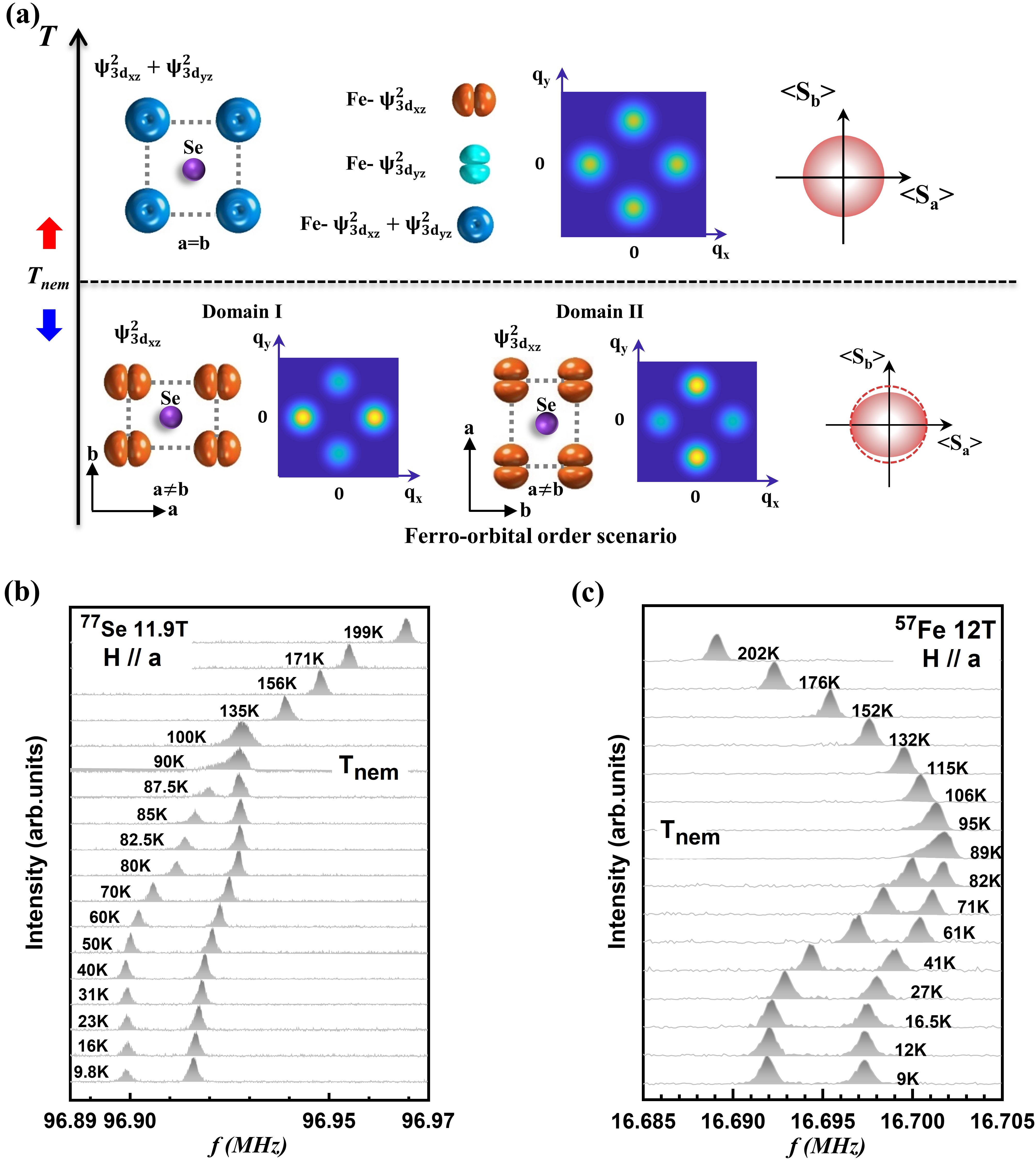}
	\caption{NMR evidence for the Ising nematic transition in FeSe. (a) Illustrative diagram of the Ising nematic transition in FeSe. Above $T_{nem}$, the system is in a paramagnetic state with an isotropic dynamic spin correlation [$ \chi{''}(q)|_{(\pi,0)} = \chi{''}(q)|_{(0,\pi)} $]. Moreover, the \dxz and \dyz  orbitals are also degenerate. In such a state, the local-spin susceptibility is also isotropic within the \textit{ab}  plane (indicated as a pink circle in the $ <S_a>$ \rm{vs} $<S_b> $ plot, in which $ <S_a> $ and $ <S_b> $ represent the local-spin polarization within the \textit{ab} plane). With the temperature lowering, a ferro-orbital ordering between the \dxz and \dyz orbitals accompanied by an orthorhombic lattice distortion (\textit{a$\neq$b}) takes place at $T_{nem}$ (indicated as the dashed black line). \hili{Two kinds of structural domains with different orbital polarizations appeare equally,} which always leads to a splitting in the NMR spectrum with \textit{H}$\parallel$\textit{a} [Figs. 1(b) and 1(c)]. Because of SOC, the local spin susceptibility is expected to gain a slight anisotropy in each domain~\cite{He2017} (indicated as a pink ellipse in $ <S_a> \textrm{vs} <S_b> $ plot). \hili{The dynamic spin correlation is coupled with the orbital order and shows an anisotropic nature.} Here, whatever the driving force is for such an Ising nematic transition, the orbital order and anisotropic dynamic spin correlation are always twisted~\cite{Fernandes2012,Fernandes2014}. (b,c) Normalized NMR spectra for \Fe and \Se nuclei with \textit{H}$\parallel$\textit{a}. Below $T_{nem}$, a clear splitting of the line shape due to the formation of the nematic domains as discussed above is shown in the spectra for both the \Fe and \Se nuclei. It should be noted that the two splitting NMR lines for the \Se nuclei show different intensities, which are caused by a partial detwinning due to unintentional strain from the tight NMR coils. We determine that such unintentional strain did not change any NMR results in the study and that the strain could be easily removed using loose NMR coils.}
	\label{fig:fig1}
\end{figure}

FeSe is a good material for exploring the above issues. Because of its prominent electronic correlation, FeSe experiences a distinctive nematic transition compared with iron-pnictide superconductors [Fig.\,\ref{fig:fig1}(a)]~\cite{Kothapalli2016,Sun2016,Wang2016}. Below the nematic transition temperature (\Tnem $ \sim $ 90 K), only orbital order (without the presence of a long-range magnetic order) is observed down to the superconducting temperature (\Tc $ \sim $ 8.9 K)~\cite{Baek2014,Bohmer2015}. Currently, the underlying mechanism for such nematic transition is still under debate~\cite{Chubukov2016}. However, a sizable band splitting due to SOC has been confirmed by recent angle-resolved photoemission spectroscopy (ARPES) measurement in FeSe. Moreover, such a SOC effect on band structure still survives in the presence of orbital order below \Tnem~\cite{Borisenko2016,Day2018}. Furthermore, a significant anisotropy for spin excitations has also been observed in the nematic state with an inelastic neutron scattering experiment (INS)~\cite{Ma2016}, suggesting a prominent role of SOC in the nematic state. \hilia{All of the above facts indicate the important role of SOC in the physical properties of the nematic state in FeSe.}

To further clarify the SOC effect and its implication in the nematic state, we conducted a site-selective nuclear magnetic resonance (NMR) study on a FeSe single crystal. In the previous NMR studies~\cite{Baek2014,Bohmer2015}, only \Se nuclei have been measurable, and the \Fe nuclei have been almost immeasurable due to their low abundance ($ \sim $2.12\%) and the small nuclear gyromagnetic ratio ( $^{57}\gamma /2\pi$ =1.376 MHz/T). By synthesizing the high-quality $^{57}$Fe-enriched ($>$98\%) FeSe single crystals, we successfully conducted NMR measurements on \Fe nuclei. \hili{The main findings from our present NMR results are summarized below. First, we reveal the predominant role of the 3\textit{d$_{xy}$} orbital on the orbital-dependent spin susceptibility, suggesting the importance of local spins for the local spin susceptibility, Second, a significant involvement of the 3\textit{d$_{xy}$} orbital in orbital reconstruction below \Tnem is determined, which unambiguously excludes a trivial ferro-orbital order in the nematic state. Third, our results also indicate a Hund's coupling-induced incoherent-to-coherent crossover on the \dxy orbital below \textit{T$_{nem}$}. Finally, due to a nontrivial SOC effect, sizable in-plane anisotropy of the spin susceptibility is observed in the nematic state, which suggests a spin-orbital-intertwined nematicity in the FeSe.}

\begin{center}
	\textbf{\uppercase\expandafter{\romannumeral2 }. METHODS}
\end{center}

The \Fe isotope-enriched ($ > $98\%) FeSe single crystalline samples were grown by a KCl-AlCl$_3$ flux method (see Supplemental Material~\cite{supple}). The resistivity and the bulk magnetic susceptibility characterizations on the \Fe isotope-enriched samples are presented in the Supplemental Material~\cite{supple}. These characterizations are the same as those of the normal samples with low abundance of \Fe isotope. The details on the NMR measurement and relevant experimental setup are also described in the Supplemental Material~\cite{supple}. Standard relaxation fitting functions and the fast Fourier transform (FFT) method were applied to analyze the NMR data. More details on the analysis of the NMR results are illustrated in the Supplemental Material~\cite{supple}.

\begin{center}
	\textbf{\uppercase\expandafter{\romannumeral3 }. RESULTS}
\end{center}

\begin{center}
	\hili{\textbf{A. The Knight shift of the \Fe nuclei: Orbital-dependent spin susceptibility}}
\end{center}

First, Figs.\,\ref{fig:fig1}(b) and \,\ref{fig:fig1} (c) show the frequency-swept NMR spectra of the \Fe and \Se nuclei for an in-plane magnetic field (\textit{H}$\parallel$\textit{a}) in the temperature range of 8--200 K. A remarkable splitting of the NMR spectra is observed below $T_{nem}$ for both nuclei, which is ascribed to the formation of two orthogonal nematic domains with \textit{H}$\parallel$\textit{a} and \textit{H}$\parallel$\textit{b} ~\cite{Wang2016,Baek2014,Bohmer2015,Fu2012,Zhou2016}. For the out-of-plane magnetic field, since both of the nematic domains are equivalent, there is no clear splitting for the NMR spectra with \textit{H}$\parallel$\textit{c} below \Tnem (see Fig. S2 in Supplemental Material~\cite{supple}). The absence of splitting in the \Fe NMR spectra with \textit{H}$\parallel$\textit{c} excludes a translational symmetry breaking in one Fe unit cell due to exotic orbital ordering~\cite{Kontani2011}, supporting the identical orbital configuration at each Fe site (see Sec. S2 of the Supplemental Material~\cite{supple}).
\begin{figure*}
	\begin{center}
		\includegraphics[width=1\textwidth]{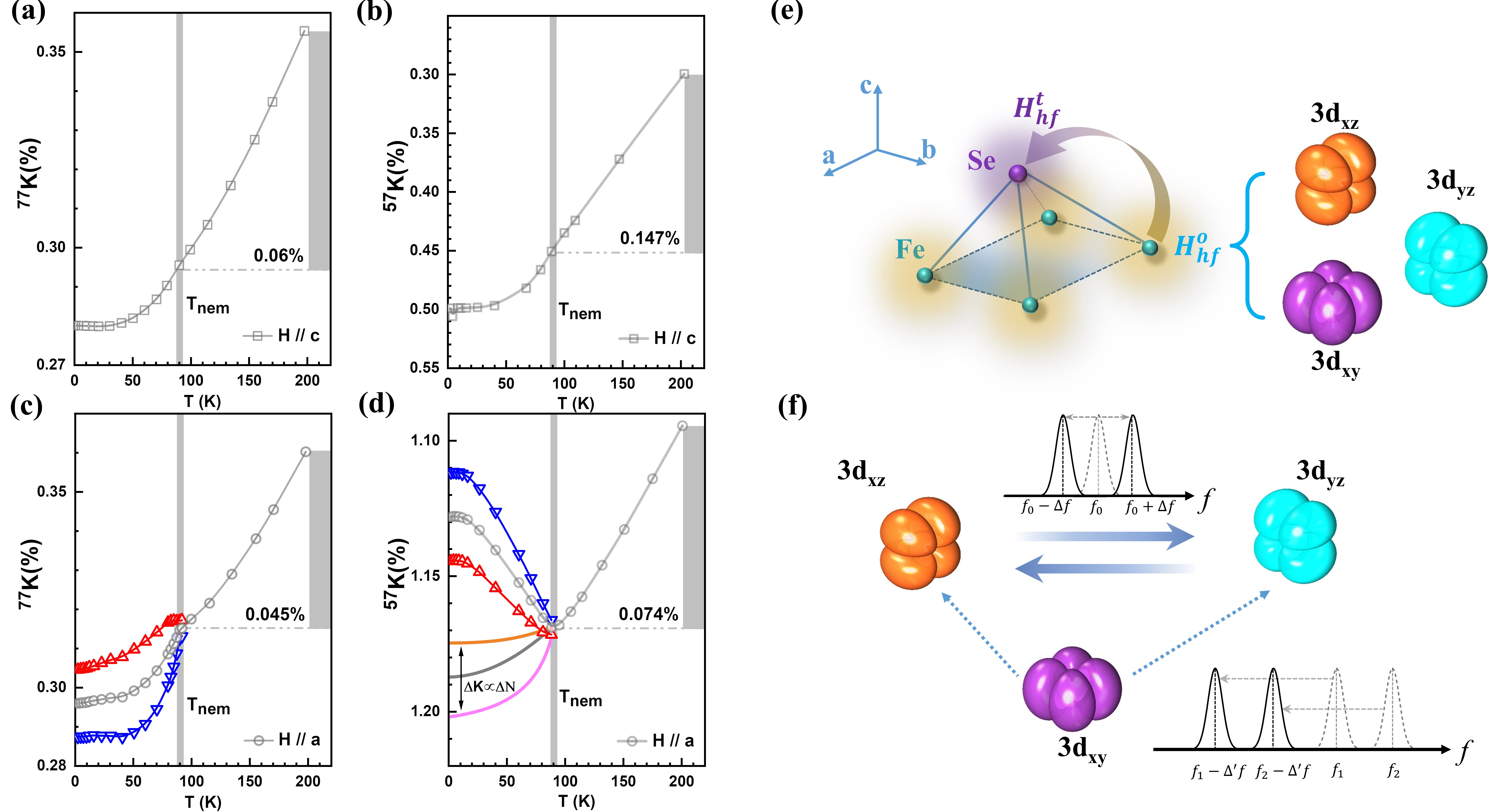}
	\end{center}
	\caption{\textrm{Knight-shift evidence for orbital-dependent spin susceptibility and orbital order beyond a trivial ferro-orbital picture.} (a)--(d) Temperature-dependent the \Fe and \Se NMR Knight shift for \textit{H}$\parallel$\textit{a} and \textit{H}$\parallel$\textit{c}. The red and blue open triangles represent the Knight shifts of the two splitting peaks below $T_{nem}$. The open gray circles below $T_{nem}$ represent the averaged value of the splitting Knight shifts. The variation of the \Fe and \Se Knight shifts above \Tnem with \textit{H}$\parallel$\textit{a} and \textit{H}$\parallel$\textit{c} are indicated as filled gray rectangles. In panel (d), based on a two-orbital model with only \dxz and \dyz orbitals, the simulated temperature-dependent \Fe Knight shifts below \Tnem\ \ are also depicted (see the Methods section in the Supplemental Material for details~\cite{supple}). The solid orange and magenta curves represent the two splitting Knight shifts. The gray bold line represents the average value of the two splitting Knight shifts. Here, $ \Delta N$ is the electron population difference between the \dxz and \dyz orbitals due to the ferro-orbital order. (e) Sketches of the primary hyperfine interactions on \Fe and \Se nuclei. The hyperfine field on \Se is transferred from the Fe 3{\it d} orbitals through the hybridization with the 4$p$ and 4$s$ orbitals at the Se sites ($ H_{hf}^t $). For the \Fe nuclei, the dominated hyperfine field is the on-site hyperfine coupling ($H_{hf}^o$) (see main text). (f) The role of the \dxy orbital on the Knight shift below \Tnem. An imbalance distribution of electrons on the \dxz and \dyz orbitals within a two-orbital model will only cause a splitting in the NMR spectra with \textit{H}$\parallel$\textit{a}. \hili{When considering the redistribution between the \dxy and \dxz($3d_{yz}$) orbitals, an overall shift to a lower resonant frequency is expected.}}
	\label{fig:fig2}
\end{figure*}

Then, we extract the Knight shift $K^\alpha$ ($\alpha$=\textit{a, b, c}) from the peak position of the NMR spectra. The temperature-dependent Knight shift for both the \Se and \Fe nuclei are shown in Fig.\,\ref{fig:fig2}. Generally, the Knight shift can be divided into two parts: $K^\alpha=K_s^\alpha+ K_{orb}^\alpha$ ($\alpha$= \textit{a, b, c}). Note that $K_s^\alpha$ is the contribution from spin d.o.f. (spin shift), which is proportional to the local spin susceptibility ($\chi_s$), and it is usually temperature dependent. $K_{orb}^\alpha$ is the contribution from the orbital d.o.f. (orbital shift), which is usually temperature-independent. By utilizing $K-\chi_{bulk}$ analysis, we can separate the spin shift and the orbital shift from the total Knight shift for both nuclei (see the Methods section of the Supplemental Material~\cite{supple}). Above \Tnem, although the bulk spin susceptibility is almost isotropic (see Fig. S3 in the Supplemental Material~\cite{supple}), the spin shift for the \Fe nuclei exhibits a sizable anisotropy between the \textit{ab} plane and out-of-plane [Figs.\,\ref{fig:fig2}(b) and \,\ref{fig:fig2}(d)]. Since $K_s^\alpha = A^\alpha \chi_s^\alpha$, where $A^\alpha$ is the hyperfine coupling tensor between the nuclear and electronic spins, the anisotropic spin shift directly reflects the anisotropy of $A^\alpha$ in this case.

Next, we discuss the origin of the anisotropic $A^\alpha$ for the \Fe nuclei. Since the transferred hyperfine coupling is usually much less anisotropic than the on-site hyperfine coupling [Fig.\,\ref{fig:fig2}(e)]~\cite{Takigawa1989}, we then consider only the on-site hyperfine coupling as the origin of the anisotropic $A^\alpha$. Given a dominant dipolar hyperfine coupling, the anisotropic part of $A^\alpha$ strongly depends on the local 3\textit{d} orbital configuration at the \Fe sites (see the Methods section in the Supplemental Material~\cite{supple}). In our case, if considering an orbital configuration with a dominant \dxz or \dyz character, we expect $ A_c < A_a $. This result is inconsistent with our experimental result. Only a dominant \dxy character can be fitted to the observed anisotropy for $ A_\alpha $. This case  indicates the predominant role of the \dxy orbital for the local-spin susceptibility. \hili{From the theoretical perspective, for simplicity, the spin susceptibility can be understood by either an itinerant spin picture or a local-spin picture. In the itinerant spin picture, the spin susceptibility is due to Pauli paramagnetism and proportional to the density of states (DOS) at Fermi level. Our NMR results indicate that if such an itinerant spin picture is applicable in FeSe, the DOS at the Fermi level should have mainly come from the \dxy orbital. However, this is in conflict with previous ARPES results, in which the primary energy band across the Fermi level have \dxz or \dyz characteristics\cite{Zhang2016,Watson2015,Watson2017,Watson2018}. The \dxy orbital has much less spectral weight at the Fermi level. This strongly suggests that the local spins need to be seriously considered together with the itinerant spins in this system. Although the Hund's coupling-induced orbital-selective electronic correlation can offer a possible mechanism for such coexistence of local and itinerant spins in FeSe\cite{Yin2011,Christensen2015,Haule2009,Fanfarillo2017,Arribi2018,You2014}, understanding the local spins is not straightforward. We will discuss more this issue in more detail later.}

In contrast to the \Fe nuclei, the spin shift for the \Se nuclei shows much less anisotropy. The hyperfine coupling tensor of the \Se is determined by the polarization of the 4\textit{p} and 4\textit{s} orbitals at the Se sites through the hybridization with the 3{\it d} orbital at the Fe sites. Only the less-extended (hybridized) Se-4{\it p} orbital can provide an anisotropic hyperfine field on the \Se. Thus, the weak anisotropy of spin shift at  the \Se sites is expected, which makes the \dxy orbital at the \Fe sites indistinguishable from the \Se NMR.

\begin{center}
	\hili{\textbf{B. Orbital reconstruction beyond a trivial ferro-orbital order}}
\end{center}

With the above knowledge on the orbital-dependent local-spin susceptibility, we next discuss the results in the nematic state. As shown in Fig.\,\ref{fig:fig2}(d), in addition to the well-known splitting of the NMR spectra due to the nematic order, the averaged in-plane Knight shift of the \Fe nuclei [$K^{ab}(T)$] also exhibits a remarkable upturn behavior below \Tnem, which is in sharp contrast to the Knight shift of the \Se nuclei. If we only consider a simple ferro-orbital order with a redistribution of the orbital population between \dxz and \dyz orbitals \hili{as suggested by previous \Se NMR results}~\cite{Baek2014}, the expected behavior of the averaged in-plane Knight shift for both \Fe and \Se nuclei should follow the same temperature dependence below \Tnem [see the caption in Fig.\,\ref{fig:fig2}(d)]. This case suggests that a simple ferro-orbital order involving only \dxz and \dyz orbitals is insufficient to understand the present \Fe NMR results. \hili{Considering previous ARPES results~\cite{Maletz2014,Zhang2016,Watson2017}, it strongly supports a necessary change of the \dxy orbital in the nematic state. For simplicity, we consider one possible change of the population on the \dxy orbital as shown in Fig.\,\ref{fig:fig2}(f). In this case, the upturn behavior of the averaged in-plane Knight shift of the \Fe nuclei is due to the population reduction on the \dxy orbital in the nematic state. However, such an explanation only makes sense within an itinerant spin picture in consideration of the Fermi surface reconstruction, which is still too simple to elucidate the underlying physics for the \Fe Knight shift below \Tnem.}

What are the underlying physics that accounted for such an change of the \dxy orbital? As we mentioned before, the orbital-dependent spin susceptibility determined by the \Fe Knight shift is beyond a simply itinerant spin picture and the local spins need to be considered as a predominant origin. If considering the Hund's coupling-induced electronic correlation in FeSCs, such coexistence of the itinerant and local spins is a natural consequence of the so-called orbital-selective electronic correlation~\cite{Haule2009,Yin2011,deMedici2009}. In FeSCs, the \dxy orbital always presents a stronger electronic correlation than the others and it should play a key role for the local spins~\cite{You2014}, which is also supported by the present \Fe NMR results. However, considering the hybridization among various orbitals, a universal incoherent-to-coherent electronic crossover is also a manifestation of the Hund's coupling-induced electronic correlation~\cite{Yin2011}, which has already been verified in FeSCs~\cite{Haule2009,Fanfarillo2017,Arribi2018}. Such electronic crossover has also been observed in most iron chalcogenide superconductors by ARPES~\cite{Yi2015}. One main feature for such electronic crossover in these iron chalcogenide superconductors is that, with the emergence of the \dxy band with decreasing temperature, a hybridization gap due to the cross of \dxy and \dxz bands opens around the M point. We propose that the change of the \dxy orbital suggested by the present \Fe Knight shift could be ascribed to a similar incoherent-to-coherent electronic crossover on the \dxy orbital. \hilia{Very recently, two ARPES experiments indicated that, since the nematic band shifting enables a band crossing between \dxy and \dxz bands at M point, a hybridization gap occurs below \Tnem~\cite{Ming2019,Huh2019} which may further promote the incoherent-to-coherent crossover of the \dxy band. Therefore, our NMR result suggests a close correlation between nematic transition and incoherent-to-coherent crossover of the \dxy band. Further theoretical calculation is needed to verify the above picture with a quantitative explanation of the results of the \Fe Knight shift.}

\begin{figure*}
 	\begin{center}
 		\includegraphics[width=1\textwidth]{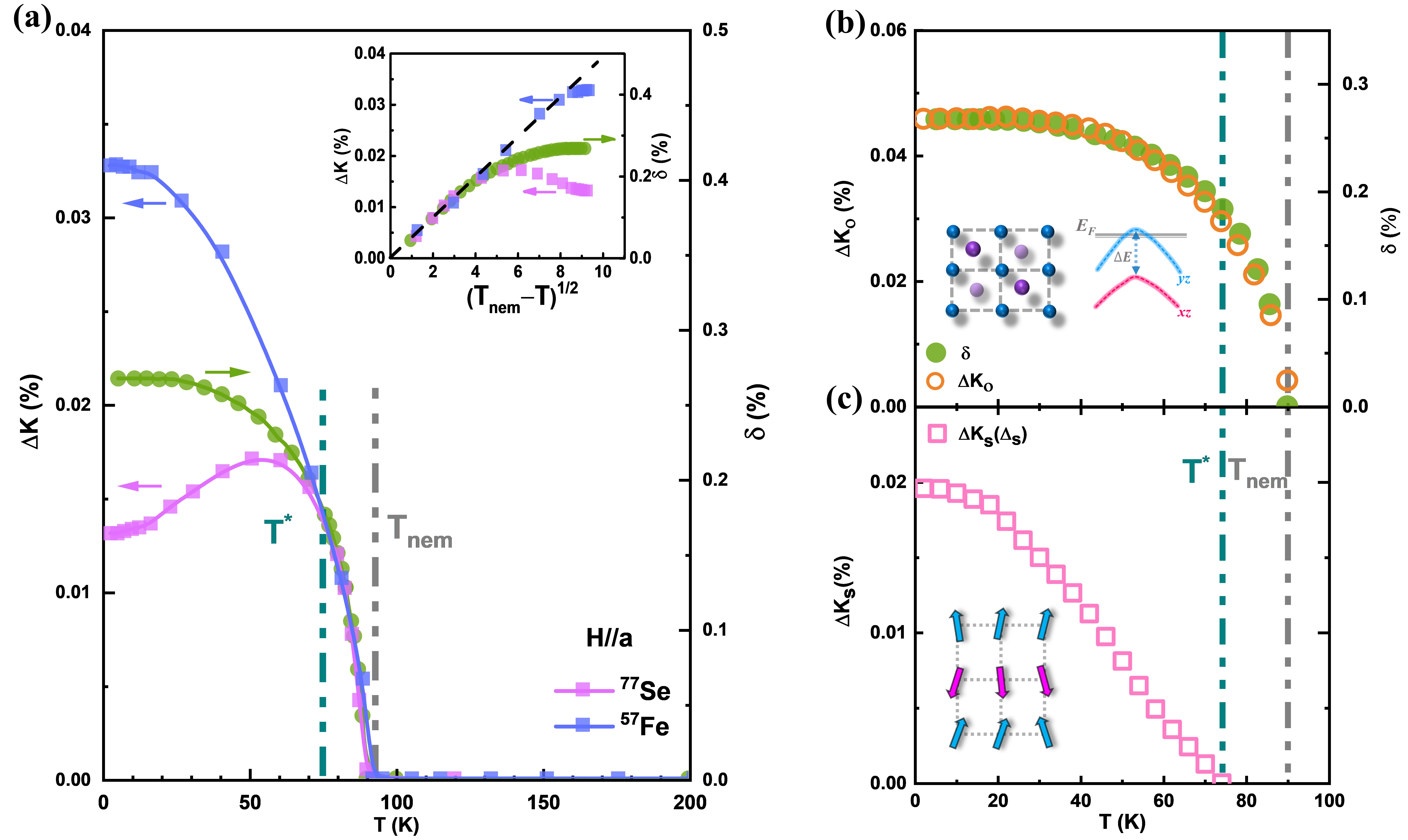}
 	\end{center}
 	\caption{Knight-shift evidence for a sizable anisotropy in spin space: temperature-dependent splitting of the Knight shift. (a)  Temperature evolution of the in-plane Knight- shift anisotropy of \Fe and \Se. The experimental data are shown as solid light-blue squares (\Fe) and solid light-magenta squares (\Se). The $ T $-dependent lattice orthorhombicity $\delta(T)(\delta=(a-b)/(a+b))$ measured through synchrotron high-energy x-ray diffraction in Ref.~\cite{Kontani2011} is also reproduced with solid dark-green circles. The inset (top right) shows the proportional relationship in the proximity of the second-order phase transition: ($ \Delta^{57}K$, $ \Delta^{77}K$, $\delta(T)$$\propto$(\Tnem-{\it T})$^{1/2}$).The discrepancy below  $T^*$ is also clearly shown. Below about 20 K, all three quantities are saturated. The solid lines are guides for the eye. (b) The temperature-dependent $ \Delta K_o(T)$ and lattice orthorhombicity $\delta(T)$. The $\Delta K_o(T)$ and $\delta(T)$ are marked as open orange circles and filled green circles. [The data are interpolated from panel (a)]. The inset shows the cartoon picture for the orbital-related band splitting and orthorhombic structural distortion. (c) The temperature-dependent $ \Delta K_s(T)$(or anisotropy of the local-spin susceptibility). Interestingly, the $ \Delta K_s(T)$ starts to arise only below $T^*$ instead of immediately below \Tnem. The inset shows a cartoon depiction of the anisotropic spin polarization in the Ising nematic state. The extracted $\Delta K_s(T)$ was interpolated from the original data shown in panel (a) with spline interpolation methods.}

	 	\label{fig:fig3}
\end{figure*}

\begin{center}
	\textbf{C. Evidence for spin-space anisotropy: Uniform spin susceptibility}
\end{center}

\begin{figure*}
	\begin{center}
		\includegraphics[width=1\textwidth]{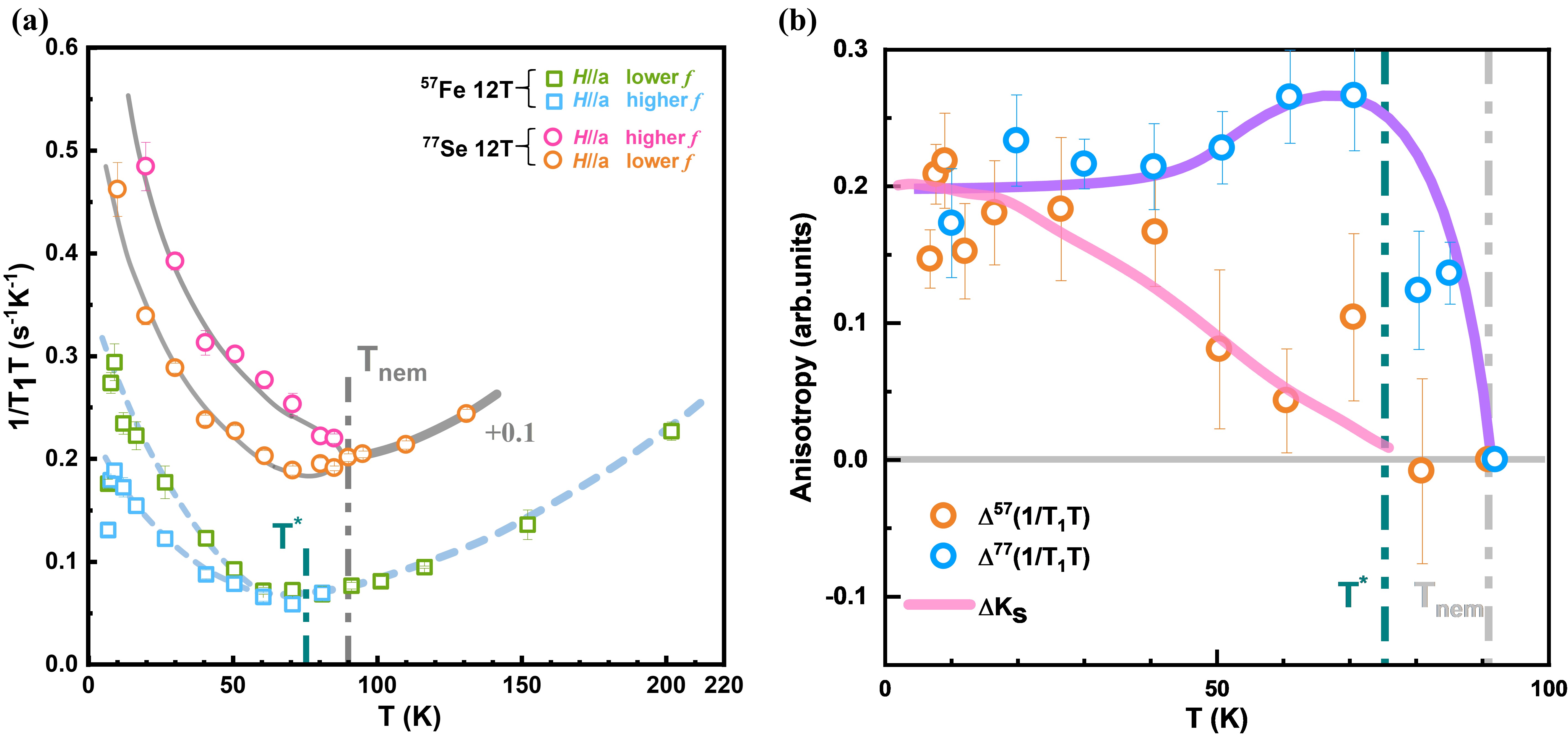}
	\end{center}
	\caption{Anisotropic nuclear spin-lattice relaxation in the nematic state: Temperature- dependent anisotropy in the spin-lattice relaxation rate divided by $T$. (a) Nuclear spin-lattice relaxation rate divided by T versus the temperature on the \Fe and \Se sites with \textit{H}$\parallel$\textit{a}. Note that $ 1/T_1T $ of \Se in different domains are marked as open magenta and orange circles. The $ 1/T_1T $ of \Fe in different domains are marked as open green and blue squares. In order to compare with \Fe and \Se, the magnitudes of the $ 1/T_1T $ for \Se were shifted up by 0.1. Interestingly, the splitting of $ 1/T_1T $ for \Fe and \Se  starts from different temperatures. (b) The temperature-dependent anisotropy in the $ 1/T_1T $ is defined as $\Delta 1/T_1T=\frac{1/T_1T|_{H\parallel a}-1/T_1T|_{H\parallel b}}{1/T_1T|_{H\parallel a}+1/T_1T|_{H\parallel b}}$. The anisotropies for both \Fe and \Se showed distinct temperature evolution below \textit{T$_{nem}$}. Note that $\Delta ^{57}1/T_1T$ is negligible between $ T^* $ and \Tnem\ and then linearly increases below $ T^* $, which is scaled with $\Delta_s(T)$. In stark contrast to $\Delta ^{57}1/T_1T$, $\Delta ^{77}1/T_1T$ shows a rapid increase immediately below \textit{T$_{nem}$} and then follows a broad peak around   65 K. Finally, $\Delta ^{77}1/T_1T$ becomes saturated with temperature below 50 K. As discussed in the main text, such a difference between \Fe and \Se is related to the detailed information on both $ A_{hf}(q) $ and $ \chi'' (q,\omega) $ (see S8 in the Supplemental Material~\cite{supple}).}
	
	\label{fig:fig4}
\end{figure*}

Next, we turn to the detailed analysis of the splitting of NMR spectrum under \textit{H}$\parallel$\textit{a} below \Tnem. As we discussed before, the splitting of the NMR spectrum is due to the formation of two orthogonal nematic domains with \textit{H}$\parallel$\textit{a} and \textit{H}$\parallel$\textit{b}. Hence, measuring the splitting of the NMR spectrum is actually equal to measuring the in-plane anisotropy of the Knight shift, which is defined as $ \Delta K(T)=K_s^a (T)-K_s^b (T)$. Since $K_s^\alpha=A_s^\alpha\chi_s^\alpha$, both of the in-plane anisotropies of the hyperfine coupling constant [$\Delta A_s(T)=^{57,77}A_s^a(T)-^{57,77}A_s^b(T))$] and the local-spin susceptibility [$\Delta\chi_s(T)=\chi_s^a(T)-\chi_s^b(T)$] contribute to $ \Delta K(T)$. When $ \Delta K(T)/(K^{ab}(T))\ll1$ and when $K^{ab}(T)$ is weakly temperature dependent below \Tnem, $ \Delta K(T)$ can be approximately expressed as a linear combination of $ \Delta A_s(T)$ and $\Delta\chi_s(T)$. A previous \Se NMR study determined that $ \Delta ^{77}K(T)$ is linearly related to the order parameter of the orbital orderig transition [$ \Delta _o(T)$]~\cite{Baek2014}. This conclusion is correct only when both $ \Delta^{77}A_s(T)$ and $\Delta\chi_s(T)$ are proportional to $ \Delta_o(T)$. \hili{Considering an orbital-driven (only the orbital d.o.f was activated) ferro-orbital order,} a finite SOC will only produce a negligible $\Delta\chi_s(T)$, which should be proportional to $ \Delta_o(T)$. This case was verified by the RPA calculation in BaFe$_2$As$_2$ by \hili{only considering a ferro-orbital order without invoking spin d.o.f.}~\cite{He2017}. In this situation, the above conjecture is naturally satisfied. \hili{This has been used as a proof to support the orbital-driven nematic order in FeSe}~\cite{Baek2014}. However, the present result proves that $\Delta\chi_s(T)$ has a sizable value and that it is not linearly coupled to $ \Delta_o(T)$.

As shown in Fig.\ref{fig:fig3}(a), when we plot the $ \Delta K(T)$ for both the \Se and \Fe nuclei together, there is a clear difference between $ \Delta ^{77}K(T)$ and $ \Delta^{57}K(T)$ below a characteristic temperature $T^* \sim $75 K. Especially for $ \Delta^{77}K(T)$, it starts to decrease below about 60 K. This behavior has already been observed in a previous NMR measurement, but its origin is still elusive~\cite{Baek2014,Bohmer2015}. A strong temperature-dependent in-plane spin susceptibility[$\overline{\chi_s^{ab}}(T) $] \hili{can give a simple explanation}~\cite{Korshunov2009}. \hili{However, such explanation seems unlikely,} because both the bulk spin susceptibility and the spin shift are almost temperature independent below 60 K (see Figs. S3 and S4 in the Supplemental Material~\cite{supple}). Furthermore, we also plot the temperature dependent orthorhombic distortion of the lattice $ \delta(T) $ in  Fig.\,\ref{fig:fig3}(a), which is proportional to $ \Delta_o(T)$~\cite{Ghosh2017}. Neither $ \Delta^{77}K(T)$ nor $ \Delta^{57}K(T)$ could be scaled to $ \delta(T) $. The key to solving the discrepancy between $\Delta^{77}K(T)$ and $ \Delta^{57}K(T)$ is to consider a nontrivial $\Delta\chi_s(T)$ that is not simply proportional to $ \Delta_o(T)$. As we discussed before, $ \Delta^{57,77}K(T)$ can be expressed as a linear combination of both $ \Delta A_s(T)$ and $\Delta\chi_s(T)$ as shown in~\cite{Baek2014}:
\begin{equation*}
\begin{aligned}
&\Delta^{57,77}K(T)\\
&= ^{57,77}A_s^{a}(T) \chi_s^{a}(T)-^{57,77}A_s^{b}(T) \chi_s^{b}(T)\\                                              &=1/2\{[^{57,77}A_s^{a}(T)-^{57,77}A_s^{b}(T)][\chi_s^a(T)+\chi_s^b(T)]+\\
&\ \ \ \  [^{57,77}A_s^a(T)+^{57,77}A_s^b(T)][\chi_s^a(T)-\chi_s^b(T)]\}\\
&=1/2\{\overline{\chi_s^{ab}}(T)\Delta_o(T)+\overline{^{57,77}A_s^{ab}}(T)\Delta \chi_s (T)\}
\end{aligned}
\end{equation*}

Because both $ \overline{\chi_s^{ab}}(T) $ and $ \overline{^{57,77}A_s^{ab}}(T) $ are weakly temperature dependent below \Tnem, the expression can be simplified as \hili{ $ \Delta^{57,77}K(T)\simeq1/2\{\overline{\chi_s^{ab}} \Delta_o(T)+\overline{^{57,77}A_s^{ab}}\Delta\chi_s(T)\}$.} \hili{Since $A_s^\alpha(T)$ can be determined by the $K_s^\alpha (T) - \chi_{bulk}(T)$ plot,} we find that the sign of the second term in the expression is different for \Se and \Fe (see Figs. S4 and S5 in the Supplemental Material~\cite{supple}). When $ \Delta\chi_s (T)$ is not simply proportional to $ \Delta_o(T)$, $\Delta K(T)$ for both of the nuclei will naturally show different temperature dependences, including the decrease in $\Delta ^{77}K(T)$. Based on the above analysis, we can separate $ \Delta_o(T)$ [Fig.\,\ref{fig:fig3}(b)] and $\Delta\chi_s(T)$ [Fig.\,\ref{fig:fig3}(c)] from $ \Delta ^{57,77}K(T)$  (see Supplemental Material~\cite{supple}). The extracted $ \Delta_o(T)$ is proportional to $ \delta(T) $ as expected. In contrast, the extracted $\Delta\chi_s(T)$ is almost negligible just below \Tnem. It starts to linearly increase only below $T ^* $$\sim$75 K and level off to a sizable value below about 20 K. A rough estimation for $\Delta\chi_s(T)/\chi_s (T)$ is about 14\% at around 10 K (see Fig. S6 in the Supplemental Material~\cite{supple}).\hili{ Such sizeable anisotropy is not consistent with a simply orbital-driven nematic order}~\cite{Baek2014,He2017}, suggesting a novel spin-orbital intertwined nematic state. This result is another main finding of the present work.

Recently, the bulk magnetic susceptibility measurements on a FeSe single crystal with a uniaxial strain also revealed similar in-plane anisotropy of the spin susceptibility~\cite{He2018}. However, the starting temperature for such in-plane anisotropy was observed slightly above \textit{T$_{nem}$}. Since our NMR experiment is performed without any external strain, the external strain effect might be a cause for such discrepancy on the starting temperature. However, the in-plane anisotropy in bulk magnetic susceptibility was ascribed to a short-range stripe magnetic order~\cite{He2018}. \hili{In this case, the short-range magnetic order will strongly broad the linewidth of NMR spectrum, which can not give a clear line splitting. This is in conflict with the present NMR experiment, in which the anisotropy of local spin susceptibility only affects the uniform spin shift rather than the linewidth} (see Sec. S7 in the Supplemental Material~\cite{supple}). Therefore, the possibility of a short-range magnetic order can be ruled out.

\begin{figure*}
	\begin{center}
		\includegraphics[width=1\textwidth]{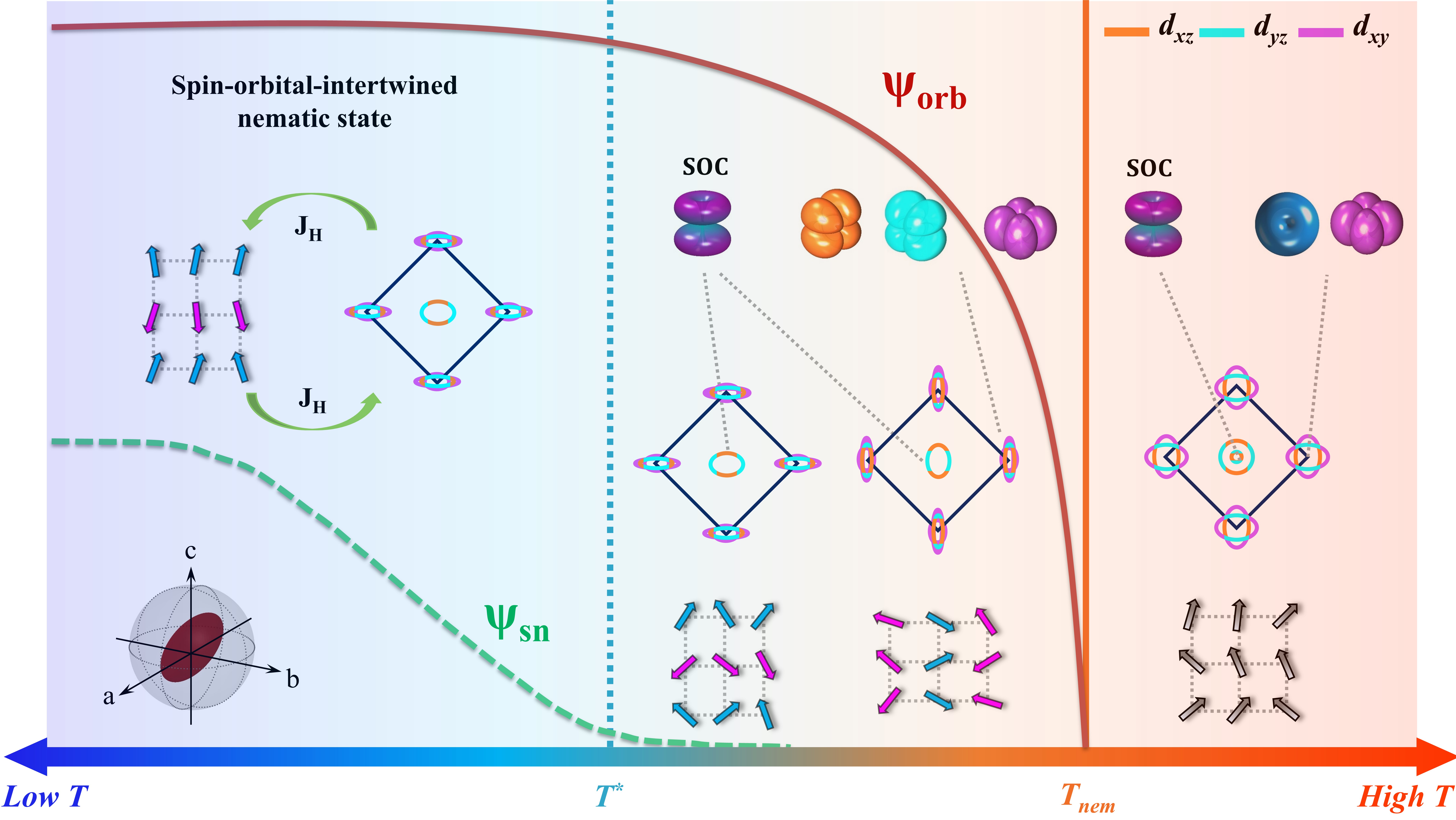}
	\end{center}
	\caption{A proposed physical scenario for the spin-orbital intertwined nematic state in FeSe. \hili{We adopt an empirical two-fluid model (detail description is presented in the main text) to explain our proposed scenario. The coherent part in this model contributed to the Fermi surface, which is related to itinerant spin d.o.f.. The incoherent part in this model stand for the local spin d.o.f. with a predominant \dxy character. These two parts are coupled by Hund's coupling. Above \Tnem, the coupling between the local and itinerant spins is incoherent. A negligible SOC effect was verified by the recent INS experiments ~\cite{Ma2016}. A nearly isotropic temperature-dependent bulk magnetic susceptibility is also confirmed, which is consistent with a negligible SOC effect (see Supplemental Material~\cite{supple}). Below \Tnem, an orbital order driven by nematic transition emerges and the corresponding temperature-dependent order parameter ($\Psi_{orb}$) obeys a standard second-order phase transition. Such orbital order break the rotational symmetry of the lattice and lead to a simultaneous Fermi surface reconstruction. Considering a finite SOC, the spin d.o.f. should also break the rotational symmetry due to orbital order. In this case, if the spin susceptibility is dominated by the local spins, and only a negligible in-plane anisotropy ($\Psi_{sn}$) can be observed through local spin susceptibility. However, a sizeable in-plane anisotropy of spin susceptibility is verified by the NMR experiment in this research (ellipsoid with a long axis along the a and c crystallographic directions at the left bottom inset), which is ascribed to an emergent coherent coupling between local and itinerant spins by Hund's coupling below \Tnem. A remarkable change of $\Psi_{sn}$ only happens below \textit{T$^*$}, which also proves that such sizeable in-plane anisotropy is related not only to the orbital order but also the the Hund's coupling-induced coherence. For more details, see the main text.}}
	\label{fig:fig5}
\end{figure*}

\begin{center}
	\textbf{D. Evidence for spin-space anisotropy: Dynamic spin susceptibility}
\end{center}

In order to further \hili{verify the spin-space anisotropy in the nematic state,} we also study the nuclear spin-lattice relaxation. In general, the quantity of $1/T_1T$   (spin-lattice relaxation rate divided by $ T $) measures the strength of low-energy spin fluctuations with $1/T_1T\sim
\Sigma_q\gamma_n^2|A_\perp(q)|^2\frac{\chi''(q,\omega)}{\omega} $, where $ \chi''(q,\omega) $ and $ A_\perp(q) $ are the $ q $-dependent dynamic spin susceptibility and the transverse hyperfine form factor, respectively~\cite{Zhou2016}. \hili{When the $ \chi''(q,\omega) $ exhibits a similar spin-space anisotropy as the uniform spin susceptibility, the $1/T_1T$ is a good indicator.} In addition to the spin-space anisotropy, the anisotropic spin excitations in $ k $ space due to stripe-type spin fluctuations [Fig.\,\ref{fig:fig1}(a)]~\cite{Fernandes2014}, \hili{which are always twisted with orbital order in the nematic state,} could also affect the anisotropy of $1/T_1T$ (see Fig. S8 in the Supplemental Material~\cite{supple}). This result has already been verified in $\rm NaFe_{1-x}Co_{x}As $~\cite{Zhou2016}. In this case, a specific $ A_\perp(q) $ is necessary to transfer the $ k $-space anisotropy into the anisotropy of $1/T_1T$. Otherwise, \hili{the anisotropic spin excitation in $ k $-space can not have a significant effect on the anisotropy of $1/T_1T$.} After careful analysis on the $ A_\perp(q) $ for both the \Se and \Fe sites, we find that only the \Se nuclei can satisfy the above requirement for $ A_\perp(q) $. Therefore, the anisotropy of $1/T_1T$ for \Se nuclei can be affected by the anisotropy in both spin and $ k $ space, but the anisotropy of $1/T_1T$ for \Fe nuclei can be affected only by the spin-space anisotropy.

As shown in Fig.\,\ref{fig:fig4}(a), the temperature-dependent $1/T_1T$ for each splitting NMR line shows a similar behavior, but with different values below \Tnem. This result means that the $1/T_1T$ for each structural domain is anisotropic within the \textit{ab} plane. If we define $\Delta 1/T_1T=\frac{1/T_1T|_{H\parallel a}-1/T_1T|_{H\parallel b}}{1/T_1T|_{H\parallel a}+1/T_1T|_{H\parallel b}}$, it is clear that the temperature-dependent $\Delta 1/T_1T$ is distinct for the \Se and \Fe nuclei [Fig.\,\ref{fig:fig4}(b)] as we expected. For the \Fe nuclei, the temperature-dependent $\Delta 1/T_1T$ only becomes significant below $ T^* $ and roughly follows the same temperature dependence as that for $ \Delta K_s(T)$, which supports a spin-space anisotropy from the dynamic spin susceptibility. In contrast, for the \Se nuclei, the temperature-dependent $\Delta 1/T_1T$ shows a nonmonotonic temperature dependence. \hili{This is because the effect of the $ k $-space and spin-space anisotropies on $\Delta 1/T_1T$ is not simply additive.} Just below \Tnem, the $\Delta 1/T_1T$ is dominated by the anisotropic spin excitations in $ k $ space, which is always twisted with the orbital order and which shows a rapid increase similar to that in $ \Delta K_o(T)$. When the temperature is lower than $ T^* $, the contribution from the spin-space anisotropy comes into play. This case leads to a slight decrease in $\Delta 1/T_1T$ for the \Se nuclei (see Fig. S8a in the Supplemental Material~\cite{supple}). In summary, the above results for the nuclear spin-lattice relaxation also confirm a significant spin-space anisotropy in the dynamic spin susceptibility, which further supports a nontrivial SOC effect in the nematic state of the FeSe.

\begin{center}
	\textbf{\uppercase\expandafter{\romannumeral4 }. DISCUSSION AND CONCLUSIONS}
\end{center}

Although the SOC always leads to a twist of spin and orbital d.o.f., the sizable spin-space anisotropy observed in the present study \hili{indicate a nontrivial SOC effect in the nematic state. In FeSCs, the driving force for a nematic transition is still under debate~\cite{Chubukov2016}. Due to the absence of a magnetic order, a simple orbital-driven nematic order has been proposed to understand the nematic transition in FeSe~\cite{Baek2014}. In this case, a trivial SOC effect will be expected with a negligible in-plane anisotropy for spin susceptibility, which should be simply scaled with the orbital order~\cite{He2017,He2018}. Clearly, this is inconsistent with our observation. However, if we consider a simply spin-driven nematic order, it would also be difficult to understand the distinct temperature-dependence of the anisotropy of spin susceptibility from the nematic order. All these facts indicate that a simple spin or orbital-driven scenario is not enough to account for these new NMR results.}

\hili{In this section, we would like to discuss the possible role of Hund's coupling in such a nontrivial SOC effect. A distinguished character for the Hund's coupling-induced electronic correlation is orbital selectivity~\cite{Yin2011,You2014,deMedici2009,Haule2009}. As discussed before, the orbital-dependent spin susceptibility determined by the \Fe NMR result confirms such orbital-selective electronic correlation and suggests a predominant role of localized spin for the local spin susceptibility. For this situation, a simplified picture for understanding the origin of spin susceptibility is based on the coexistence of both local and itinerant spins. Although such a picture is quite practical for explaining the results of the \Fe Knight shift, it also highlights another fundamental issue for FeSCs, which is how to understand the origin of localized spins in FeSCs. Answering this question is beyond the scope of the present work. Hence, we only introduce a promising approach that is based on an empirical two-fluid model for FeSCs~\cite{Yang2010,You2011,You2013,You2014}, to discuss this issue. In this model, the total electronic spectral weight is separated into two distinct parts. One part is relatively coherent and it is related to the itinerant spins. The other part is relatively incoherent and it is related to the local spins. The intra-orbital Hund's coupling leads to an on-site ferromagnetic coupling between these itinerant spins and the localized spins. Based on above empirical model, we propose a simple picture to understand the above nontrivial SOC effect in the nematic state. \hilia{As shown in Fig.\,\ref{fig:fig5}, as the incoherent-to-coherent crossover on the \dxy orbtial is developing below \Tnem, a coherent on-site ferromagnetic coupling between localized spins and itinerant spins is also simultaneously built up.} This means that the electronic system evolves from an incoherent state with two spin fluids into a coherent state with a single spin fluid. In such coherent state, although the orthorhombicity due to orbital order is less than 0.3 percent, the SOC induces a strong coupling between spin and orbital d.o.f., which gives rise to sizeable in-plane anisotropy of the local spin susceptibility. This is ascribed to a nontrivial SOC effect due to the Hund's coupling-induced electronic correlation, which is also the manifestation of a spin-orbital-intertwined nematic state.}

Following the above scenario, the superconducting state may also be strongly affected by SOC. Recently, an extremely anisotropic superconducting gap has been observed in bulk FeSe~\cite{Kushnirenko2018,Liu2018}. Moreover, the superconducting pairing has also been revealed as orbital selective~\cite{Sprau2017}. All of these experimental results may have a strong link with the spin-orbital-intertwined nematicity observed in the present work. Interestingly, a previous NMR study found that the Knight shift stayed almost constant across the superconducting transition temperature~\cite{Baek2014,Bohmer2015}, which leads to an argument for \textit{p}-wave pairing~\cite{Liu2018}. This case is worthy of further exploration by considering a nontrivial SOC effect. In addition, another possible SOC effect on the superconducting state has also been suggested by a recent high-field study~\cite{Kasahara2014}. Currently, the SOC effect on the superconducting state is still elusive, which calls for further studies.

In conclusion, by utilizing a site-selective NMR measurement, the Hund's coupling-induced orbital-selective electronic correlation is reconfirmed by investigating the orbital-dependent local-spin susceptibility in FeSe, suggesting a predominant role for the \dxy orbital in spin susceptibility. This result also indicates that the itinerant spins alone were not enough to account for spin susceptibility and that local spins should be seriously considered as a predominant origin. Furthermore, the temperature evolution of the orbital-dependent local-spin susceptibility reveals a remarkable change of the \dxy orbital in the nematic state, which is definitely beyond a trivial ferro-orbital order picture. \hilia{This result also enables us to propose a Hund's coupling-induced incoherent-to-coherent crossover on the \dxy orbital, which may have a close correlation with a recently discovered hybridization gap around the M point. Finally, due to a nontrivial SOC effect, sizeable in-plane anisotropy of the spin susceptibility is observed in the nematic state, which suggests a spin-orbital-intertwined nematicity rather than simply spin- or orbital-driven nematicity in the FeSe.} \\

\begin{center}
	\textbf{ACKNOWLEDGMENTS}
\end{center}

We thank Y. Li, F. Wang, R. Yu, G. Chen, K. Jin, X. L. Dong, R. Zhou, M. -H. Julien, R. M. Fernandes, J, Schmalian and Z. Y. Weng for stimulating discussions. This work was supported by the National Natural Science Foundation of China (Grants No. 11888101 and No. 11522434), the National Key R\&D Program of the MOST of China (Grants No. 2016YFA0300201 and No. 2017YFA0303000), the Strategic Priority Research Program of the Chinese Academy of Sciences (Grant No. XDB25000000), and the Anhui Initiative in Quantum Information Technologies (Grant No. AHY160000).\\

\end{document}